\documentclass[superscriptaddress,aps,prl,twocolumn,showpacs,preprintnumbers,amsmath,amssymb,floatfix]{revtex4}
\usepackage[T1]{fontenc}
\usepackage{graphicx}
\usepackage{times}
\newcommand{\Tc}{$T_{\mathrm{c}}$}
\newcommand{\Tg}{$T_{\mathrm{g}}$}
\newcommand{\Tx}{$T_{\mathrm{x}}$}
\newcommand{\ud}{\mathrm{d}}
\begin{document}
\title{Anomaly of the non-ergodicity parameter and crossover to white noise\\ in the fast relaxation spectrum of a simple glass former}
\author{S.\,V.~Adichtchev} 
\author{St.~Benkhof} 
\author{Th.~Blochowicz} 
\author{V.\,N.~Novikov} 
\author{E.~R\"ossler} 
\altaffiliation{corresponding author}
\author{Ch.~Tschirwitz} 
\affiliation{Physikalisches Institut EP~II, Universit\"at Bayreuth, 95440 Bayreuth, Germany}
\author{J.~Wiedersich}
\affiliation{Physikalisches Institut EP~II, Universit\"at Bayreuth, 95440 Bayreuth, Germany}
\affiliation{Technische Universit\"at M\"unchen, Lehrstuhl f\"ur Physik Weihenstephan, V\"ottinger Str.~40, 85350 Freising, Germany}
\date{November 14, 2001}
   
\begin{abstract}
We present quasi-elastic light scattering (LS)
and dielectric (DS) spectra of the glass former $\alpha$-picoline 
($13\,\mathrm{K} < T < 320\,\mathrm{K}$). 
At high temperatures the evolution of the susceptibility minimum is well described by mode coupling theory (MCT) 
 yielding a critical temperature $T_{\mathrm{c}} = 162\,\mathrm{K} \pm 5\,\mathrm{K}$. 
At $T_{\mathrm{c}} > T \geq T_{\mathrm{g}}$
 the excess wing of the $\alpha$-process identified in the DS spectra and characterized by a power-law exponent $c$ is rediscovered in the LS spectra. 
Introducing a universal evolution for $c = c(\lg \tau_{\alpha})$ as suggested by DS data of several glass formers, 
the fast dynamics spectrum is singled out, allowing to determine the non-ergodicity parameter $f(T)$. 
The latter shows the predicted cusp-like anomaly 
identifying $T_{\mathrm{c}}$ as well. 
Another discontinuous change of $f(T)$ is observed at $T_{\mathrm{g}}$. 
The fast dynamics spectra exhibit a crossover to ``white noise'' ($T < T_{\mathrm{c}}$). 
Concerning the fast dynamics, we conclude that idealized MCT predictions hold even below $T_{\mathrm{c}}$.
\end{abstract}
\maketitle

The molecular slowing-down in glass forming liquids has been investigated by neutron (NS) and light scattering (LS) as well as by dielectric spectroscopy (DS) \cite{ref1}, and the predictions of mode coupling theory (MCT) have been found to hold above the critical temperature \Tc{}, the latter being about 20\,\% above the glass transition temperature \Tg. 
Below \Tc{} the situation is less clear. 
Concerning the temperature dependence of the non-ergodicity parameter $f(T)$, 
which is expected to show a cusp-like behavior, 
agreement is less convincing and a matter of debate \cite{ref1,ref2}. 
Related to this is the prediction of a ``knee'' in the fast dynamics spectrum, 
i.\,e.\ a power law behavior, $\chi''(\nu) \propto \nu^a$, 
with an exponent $a$ changing from $a \approx 0.3$ at high frequencies to $a = 1.0$ at low frequencies.
A crossover to a ``white noise'' spectral density is predicted upon cooling towards $T < T_{\mathrm{c}}$. 
However, up to now no such behavior has been confirmed. 
The frequency window of, e.\,g., LS techniques is restricted to frequencies $\nu \gtrsim 0.5$\,GHz, 
and one could argue that the knee cannot be identified unambiguously. 
Recent MCT analyses indeed have shown that the experimental data are compatible with the presence of a knee \cite{ref3}. 
In addition, below \Tc{} the analysis is hampered by the fact that the time constant $\tau_\alpha$ 
of the $\alpha$-process actually does not diverge at \Tc{} as presumed by idealized MCT. 
Hence, contributions from the $\alpha$-process have to be taken into account to 
evaluate the fast dynamics spectrum. 
However, there is no generally accepted way to do this. 

The aim of this contribution is to perform an analysis of the susceptibility spectra 
of a simple glass former as obtained by 
tandem Fabry-Perot interferometry (TFPI) and by DS 
in the temperature range 2.5\,\Tg{} $>T>$ 0.1\,\Tg. 
Most LS measurements reported so far do not include the temperature range below \Tg. 
We carefully apply additional interference filters in order to suppress contributions from higher orders in the TFPI spectra
, as is required to obtain realiable LS spectra \cite{ref5,ref6,ref7}.
In particular, we address the question of how the susceptibility minimum evolves below \Tc. 
We will demonstrate that the excess wing of the $\alpha$-process, 
identified in dielectric experiments, is also observed in LS spectra. 

We have studied the glass former $\alpha$-picoline (cf.\,Fig.~\ref{fig2})
\begin{figure}
\includegraphics*[width=.42\textwidth]{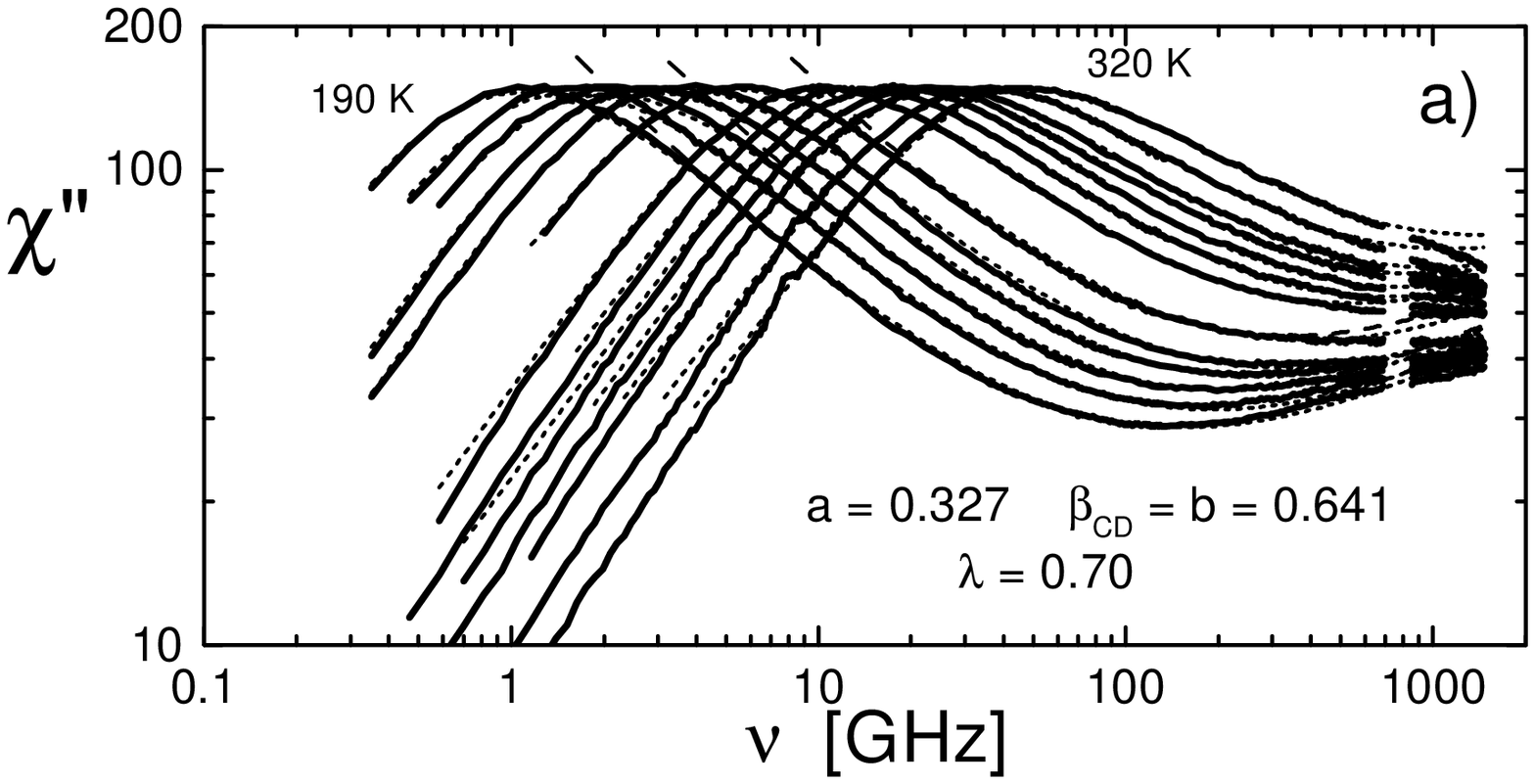}\\
\includegraphics*[width=.42\textwidth]{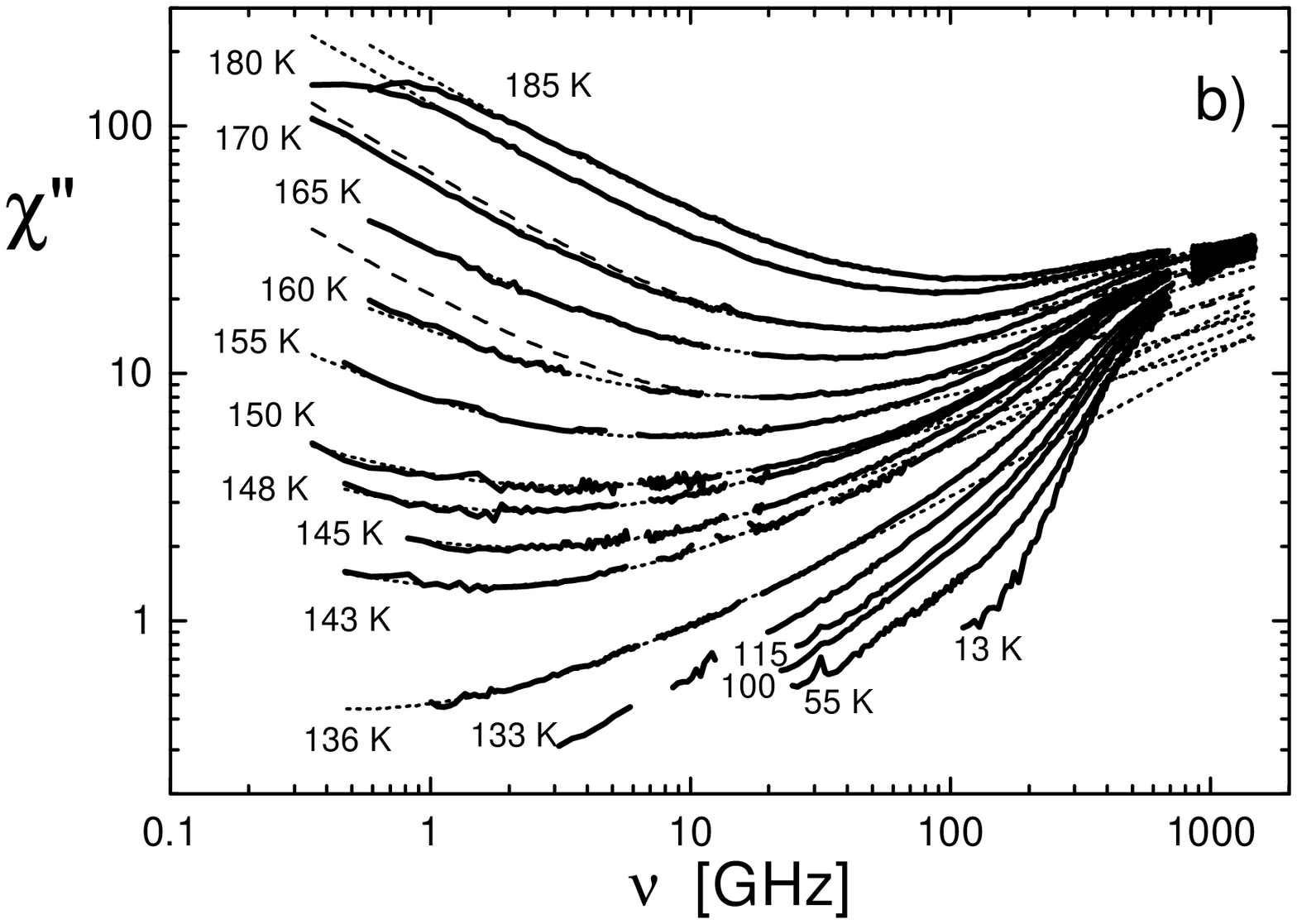}\\
\includegraphics*[width=.42\textwidth]{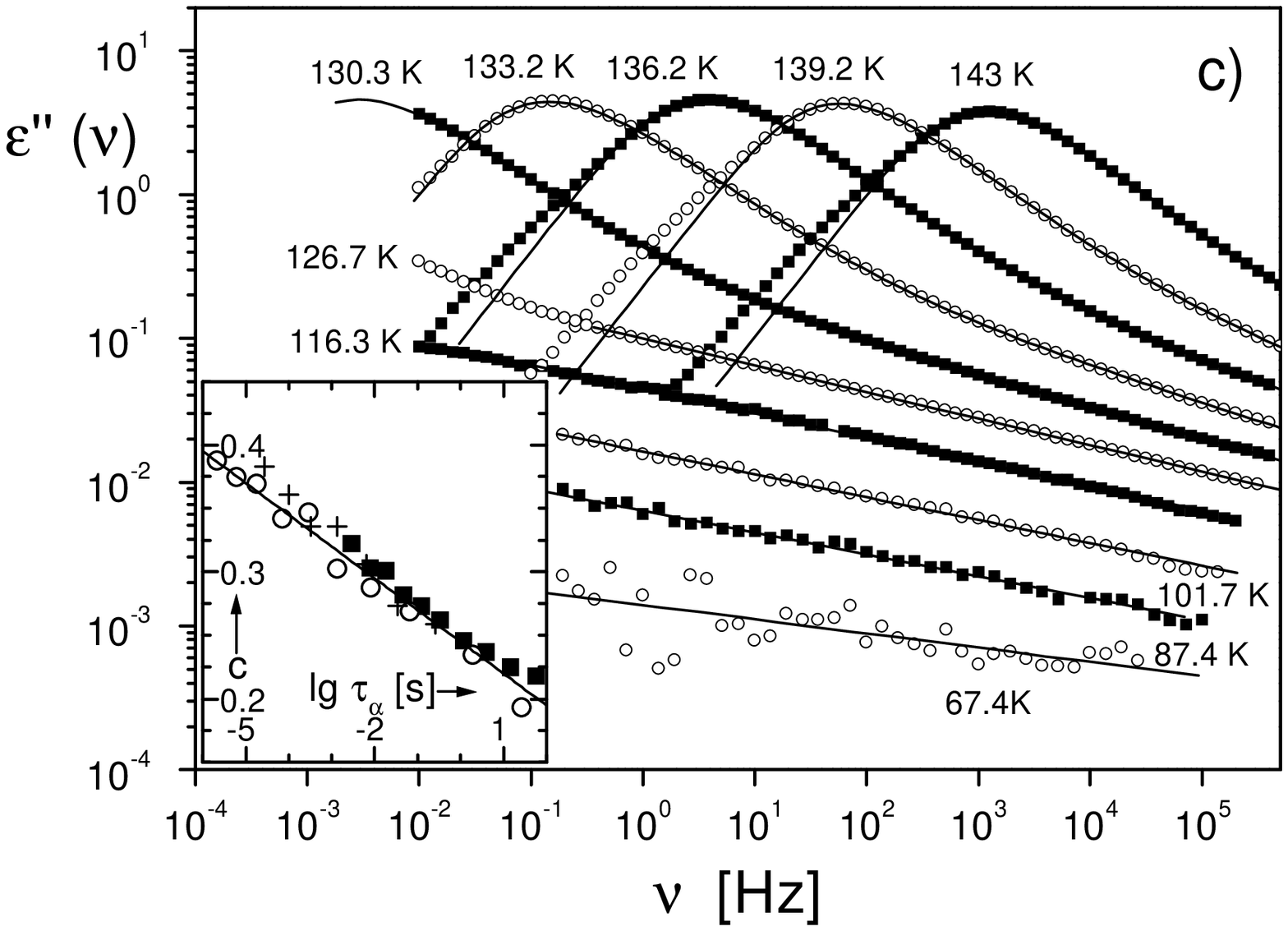}
\caption{a) Light scattering (LS) spectra of picoline above 
185\,K (190\,K, 195\,K, 200\,K, 205\,K, 210\,K, 220\,K, 250\,K, 270\,K, 280\,K, 300\,K, 320\,K, solid lines), 
fit of the minimum by MCT, eq.~(\ref{eq1}) (dashed lines), 
fit by the phenomenological approach (dotted lines); 
b) LS spectra below 185\,K (solid lines), 
fit by the superposition of two power-laws (dotted lines), 
MCT minimum curve (dashed lines); 
c) spectra from dielectric spectroscopy (symbols), 
fit interpolating $\alpha$-relaxation peak and excess wing (solid lines); 
insert: exponent $c$ of the excess wing as a function of $\lg(\tau_\alpha)$ 
for picoline (squares), glycerol (circles) and propylene carbonate (crosses),
solid line: linear fit. 
\label{fig1}}
\end{figure}
which has the same molecular weight and a similar structure as toluene but a significantly higher dipole moment. 
Picoline has a smaller tendency to crystallize and allows us to compile more LS data at 
\Tg{} $< T <$ \Tc{} \cite{ref8}. 
By DSC \Tg{} is found to be at 133\,K.

Let us sum up some predictions of idealized MCT \cite{ref1}. 
The fast dynamics and the $\alpha$-process are separated by a minimum in the suszeptibility at 
$T >$ \Tc{}, which can be interpolated by 
\begin{equation}
\chi'' (\nu) =\chi''_{\min}
\left[
\frac{%
b\left(\nu / \nu_{\mathrm{min}}\right)^{a}+
a\left(\nu / \nu_{\mathrm{min}}\right)^{-b}%
}{%
a + b
}
\right],
\label{eq1}
\end{equation}
where $\nu_{\mathrm{min}}$ and $\chi''_{\mathrm{min}}$ are the frequency and the amplitude of the minimum. 
The exponents $a$ and $b$ describe 
the low-frequency behavior of the fast dynamics and the high-frequency part of the $\alpha$-process, respectively; 
they are interrelated via the exponent parameter $\lambda$ \cite{ref1}. 
The temperature dependences are given by 
$\chi''_{\mathrm{min}} \propto \sigma^{1/2}$, $\nu_{\mathrm{min}}\propto \sigma^{1/2a}$
for the amplitude and the frequency of the minimum 
and $\tau_{\alpha} \propto \sigma^{-\gamma}$ for the time scale of the $\alpha$-process, 
where $\sigma = \frac{T-T_{\mathrm{c}}}{T_{\mathrm{c}}}$.
The exponent $\gamma$ is related to the exponents 
$a$ and $b$ 
via $\gamma=\frac{1}{2a}+\frac{1}{2b}$.
The critical temperature \Tc{} can be identified by the temperature dependence of 
$\chi''_{\mathrm{min}} (T)$, $\nu_{\mathrm{min}}(T)$ and $\tau_\alpha(T)$. 

Below \Tc{} the fast relaxation spectrum is expected to exhibit 
a crossover from a power-law behavior with exponent $a$ at high frequencies/temperatures 
to a white noise spectrum ($a = 1$) at low frequencies/temperatures. 
As a consequence of the appearance of this ``knee'' a decrease of the relaxation strength is expected upon cooling 
and a characteristic temperature dependence is predicted for the non-ergodicity parameter $f(T)$
\begin{equation}
f(T) = f_{\mathrm{c}}, \quad T>T_{\mathrm{c}}; 
\quad f(T) = f_{\mathrm{c}} + h |\sigma |^{1/2}, \quad T<T_{\mathrm{c}}.
\label{eq2}
\end{equation}
The quantity $1-f$ presents the fraction which decays due to processes faster than the $\alpha$-process and is given by the integral over the susceptibility of the fast relaxation $\chi''_{\mathrm{fast}}$,
$
1-f \propto \int_{-\infty}^{\infty} \chi''_{\mathrm{fast}} (\nu)\ \ud \ln\nu.
$

Fig.~\ref{fig1} shows the susceptibility spectra as obtained by LS (Fig.~\ref{fig1}a, b) 
and by DS (Fig.~\ref{fig1}c). 
Picoline is a type A glass former \cite{ref9}, since no slow $\beta$-process can be identified in the DS data. 
Rather a ``wing'' at high frequencies shows up, 
which degenerates to a simple power law contribution with an exponent 
$c \simeq -0.1$ at $T < $ \Tg{}. 
Fig.~\ref{fig1}a also displays MCT interpolations by Eq.~\ref{eq1} of the susceptibility minimum 
with an exponent parameter $\lambda = 0.70$ fixing $a = 0.327$ and $b = 0.641$ (dashed lines) \cite{ref1}. 
Clearly, at high temperatures the interpolation well reproduces the minimum, 
whereas at low temperatures the agreement becomes worse (cf.\ Fig.~\ref{fig1}b). 
The parameters $\chi''_{\mathrm{min}}$, $\nu_\mathrm{min}$ and $\tau_\alpha$ (cf.\ below) 
all exhibit the behavior expected from MCT at $T > $ \Tc{} (cf.\,Fig.~\ref{fig2}): Extrapolating the scaling laws, one obtains consistently \Tc{} $= 162\,\mathrm{K} \pm 5\,\mathrm{K}$. 
\begin{figure}
\includegraphics*[width=.3\textwidth]{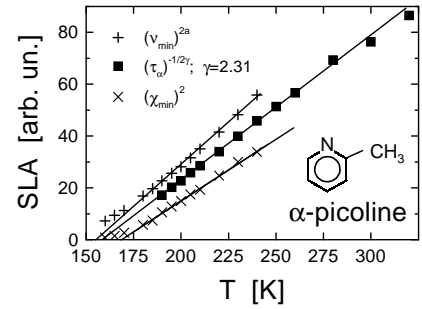}
\caption{Linearized scaling law amplitudes (SLA) obtained from
$\chi''_{\min} (T)$, $\nu_{\min} (T)$ and $\tau_{\alpha} (T)$ (cf.\ text); a = 0.327, b = 0.641, $\gamma$ = 2.31; 
solid lines: linear fits
\label{fig2}}
\end{figure}
Thus, idealized MCT provides a very satisfying interpolation at high temperatures up to the fluid regime. 
However, one observes some deviations from the scaling laws at temperatures below some  180\,K (cf.\,Fig.~\ref{fig2}). 

\begin{figure}
\includegraphics*[width=.42\textwidth]{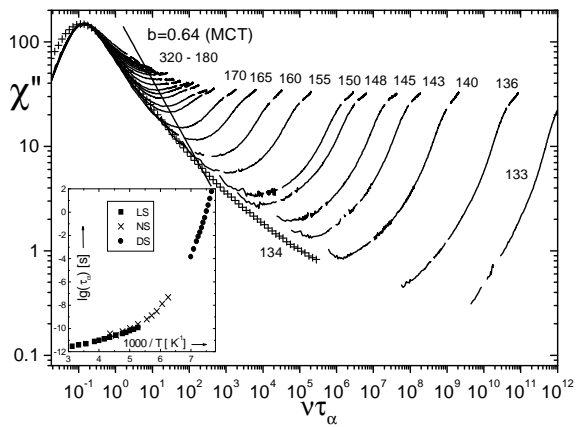}
\caption{Light scattering (LS) spectra plotted as a function of reduced frequency $\nu\tau_\alpha$. 
Solid line: MCT prediction for the position of the minimum; 
crosses: dielectric spectrum at $T= 134$\,K.
Numbers indicate temperatures in $[\mathrm{K}]$.
Insert: temperature dependence of the time constant  $\tau_\alpha$ as obtained by 
LS,  NS and DS.
\label{fig3}}
\end{figure}

A full description of the relaxation spectrum, including the $\alpha$-relaxation peak, is not given by 
idealized MCT. 
Hence, in order to extract the time constant $\tau_{\alpha}$ we choose a phenomenological approach,
assuming an additive superposition of a 
Cole-Davidson,  $
\chi''_{\mathrm{CD}} (\nu) \propto \Im (1+ \mathrm{i}\omega\tau)^{-b}
$, 
and a power-law, $\nu^{a'}$, susceptibility \cite{ref8}
for $\alpha$-process and fast dynamics, respectively. 
The fits for $180\,\mathrm{K} <T < 250\,\mathrm{K}$ 
are practically indistinguishable from those of MCT around the minimum
(here the exponents are kept the same as in the MCT analysis, $a'=a$), 
but now include the full relaxation spectrum (cf.\,Fig.~\ref{fig1}a, dotted line). 
The time constants $\tau_{\alpha}$ are plotted in Fig.~\ref{fig3} (insert). 
Including those from NS \cite{ref10} and DS the full temperature range of the supercooled liquid is covered. 
Comparing the time constants $\tau_{\alpha}$ from the different techniques similar results are observed. 

To unravel the origin of the deviations from MCT's scaling laws at low temperatures 
(cf.~Fig.~\ref{fig2}, $T < 180$\,K) 
we present in Fig.~\ref{fig3} the LS data in a way to provide a master plot for the $\alpha$-relaxation peak, 
i.\,e., $\chi''$ is plotted as a function of $\nu\tau_{\alpha}$, 
where $\tau_{\alpha}$  is taken from interpolating $\tau_{\alpha} (T)$.
According to MCT, the loci of the minimum are expected to obey the relationship
$\chi''_{\min} (\nu) \propto (\nu_{\min} \tau_\alpha)^{-b}$. 
This is indeed observed for $T > 180$\,K. 
Below this crossover temperature subsequently called \Tx{}  deviations show up. 
As also suggested by Fig.~\ref{fig2}, 
the minima are shifted to higher frequencies $\nu_{\min}$ and to higher amplitudes $\chi''_{\min}$ 
as compared to MCT predictions. 
It appears that the minimum is determined not any longer by the 
high-frequency power-law of the $\alpha$-process (with exponent $b$) 
but rather by some \emph{additional} spectral feature 
which only appears below $T_{\mathrm{x}} \simeq  180$\,K. 
Since picoline does not exhibit a slow $\beta$-process, 
the only candidate for a change in the susceptibility 
is the excess wing, which is clearly seen in the DS spectra (cf.\,Fig.~\ref{fig1}c). 
From DS data of glycerol and propylene carbonate \cite{ref9,ref11}, 
which have been measured to higher frequencies, 
it is seen that the excess wing disappears at high temperatures. 
We conclude that 
the deviations of $\chi''(\nu)$ from the MCT description 
are due to the appearence of the excess wing below a certain temperature. 

The excess wing is observed by DS in all supercooled liquids 
provided that it is not obscured by the presence of a slow $\beta$-process. 
Its very similar spectral shape has been stressed by several authors 
and may be described by a power-law contribution with exponent $c < \beta_{\mathrm{CD}}$ \cite{ref9,ref12}. 
In order to single out the fast dynamics contribution it is important to quantify $c(T)$. 
Recently, we have proposed a distribution of correlation times 
which allows to fit almost perfectly the spectral shape of both the $\alpha$-process peak and the wing and to extract $c$ \cite{ref9}. 
In Fig.~\ref{fig1}c we have included such fits. 
Moreover, analyzing several glass formers we find that $c$ shows a universal behaviour when plotted as function of $\lg(\tau_{\alpha})$ (Fig.~\ref{fig1}c, insert). 
Contrary, the parameter $\beta_{\mathrm{CD}}$ 
describing the width of the $\alpha$-peak is not universal. 

These are results obtained from DS, 
and a priori there are no arguments that this universality also holds for the LS data. 
However, recent LS measurements of glycerol clearly show 
that the wing manifests itself in a very similar way as in DS \cite{ref13}. 
Moreover, a direct comparison of DS and LS data in Fig.~\ref{fig3} demonstrates 
that the excess wing is at least similar for both methods. 
Thus, let's assume that $c(T)$ is the same in both DS and LS. 
We further assume that the susceptibility minimum below 180\,K is described 
by a sum of two power law contributions, 
$\chi''(\nu) \propto A \nu^{-c} + B \nu^{a'}$
(cf.~dotted-lines-fit in Fig.~\ref{fig1}b), yielding an excellent agreement with the data. 
The exponent $a'$ is the exponent attributed to the fast dynamics below \Tx{}. 
Inspecting Fig.~\ref{fig1}b it is obvious that $a' \neq a$ 
at lower temperatures, 
e.\,g., $a' = 0.72$ is found at 115\,K. 
For this low temperature ($T<T_{\mathrm{g}}$), $a'$ can directly be read from the data 
since no interference with the wing contribution is expected. 
Thus, it becomes obvious that $a'$ strongly depends on temperature 
in the interval \Tg{} $< T <$ \Tx{}.

\begin{figure}
\includegraphics*[width=.42\textwidth]{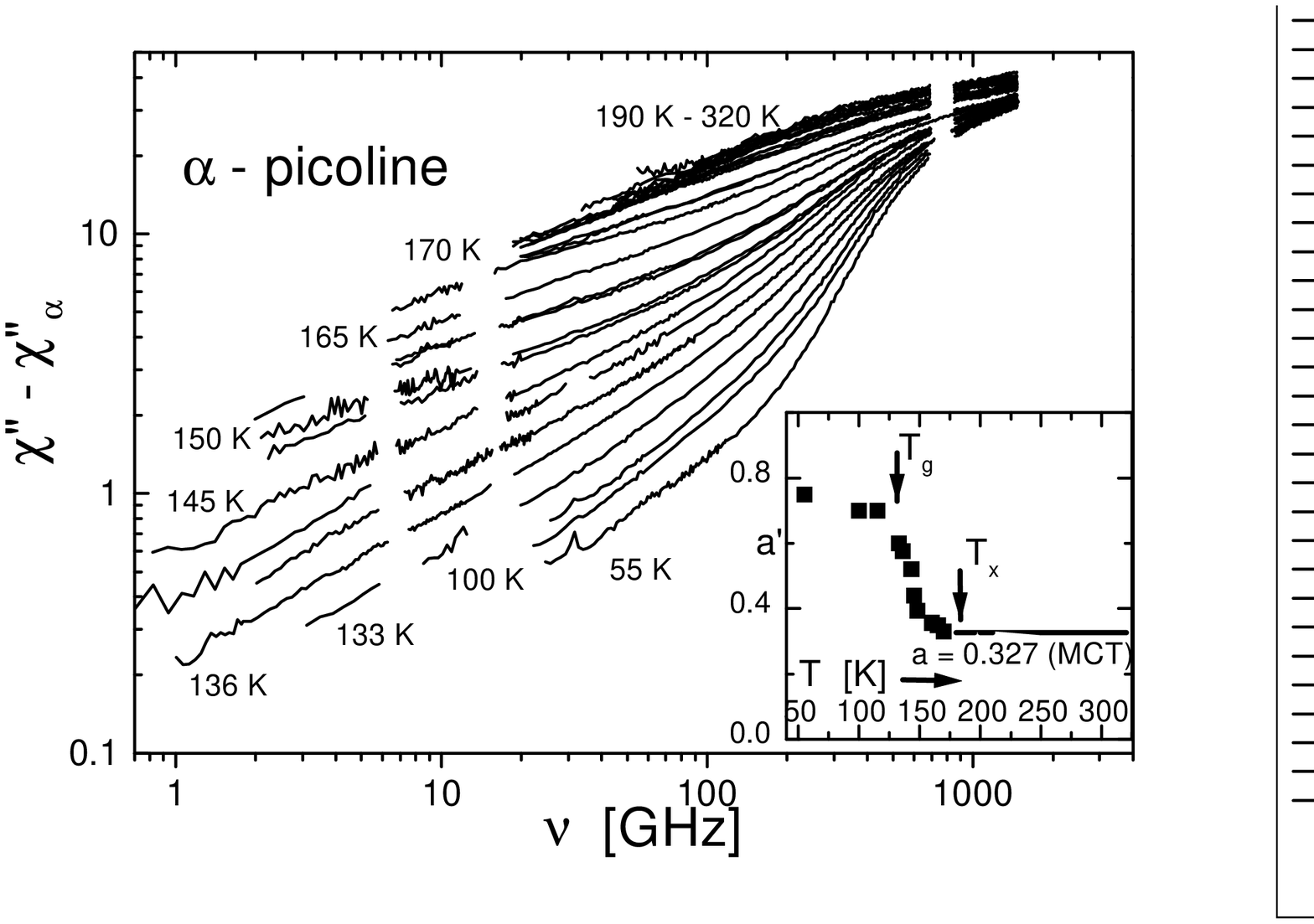}
\caption{Fast dynamics spectrum, $\chi''_{\mathrm{fast}}(\nu) = \chi''(\nu) - \chi''_\alpha (\nu)$; 
insert: temperature dependence of the exponent $a'$.
\label{fig4}}
\end{figure}

Fig.~\ref{fig4} displays the LS spectra 
after subtracting the contribution of the $\alpha$-process including the excess wing. 
In agreement with the phenomenological fit in Fig.~\ref{fig1}a 
the power-law of the fast dynamics with a constant amplitude and an exponent $a = 0.327$ 
is rediscovered at $T > $ \Tx{}. 
Referring to MCT this is the critical spectrum. 
Below \Tx{} $\simeq 180$\,K the spectra can still be described by power-laws 
but first their amplitude decreases and 
on further cooling 
the exponent $a'$ increases as well.  
The temperature dependence of $a'$ is plotted in Fig.~\ref{fig4} (insert). 
It is essentially constant at $T <$ \Tg{}, then drops above \Tg{} 
and finally reaches the level of $a = 0.327$ close to \Tx{}. 
Below \Tx{} the exponent of the fast dynamics increases towards one;
however, this trend stops at \Tg{}.

\begin{figure}
\includegraphics*[width=.3\textwidth]{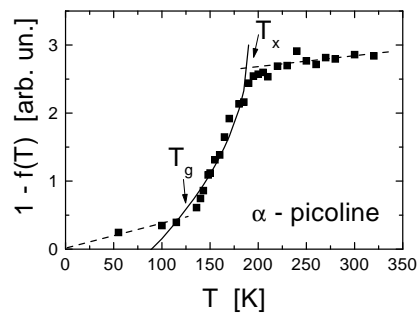}
\caption{Non-ergodicity parameter $f$: plotted is $1 - f(T)$;
solid line: square root law
(eq.~\ref{eq2}); dashed lines: guide to the eye.
\label{fig5}}
\end{figure}

Having singled out the fast dynamics contribution we are able to determine the non-ergodicity parameter $f$. 
More precisely, we integrate the fast dynamics spectra 
and get a quantity which is proportional to $1 - f$ (cf.\,above).
The integration is performed in the frequency interval 0 -- 200\,GHz. 
Fig.~\ref{fig5} displays the results.  
Obviously, the anomaly expected by MCT is discovered: 
Above 180\,K only a weak temperature dependence of $1- f$ is discernible; 
below a strong decrease is observed with decreasing temperature. 
This decrease can be fitted by the square root law 
(cf.\,Eq.~2). 
At $T \simeq$ \Tg{} the strong temperature dependence halts and 
only a weaker one persists at  $T <$ \Tg{}.

This contribution is aimed at 
describing the evolution of the LS susceptibility on cooling a glass former 
from high temperatures ($T \gg $ \Tx{}) to temperatures in the glassy state ($T \ll$ \Tg{}). 
Since no 
simple and generally accepted 
predictions exist for the low temperature regime, $T <$ \Tc{}, 
and since the  frequency window is limited, 
our analysis is phenomenological in nature and has to rely on certain assumptions. 
Nevertheless, 
a clear scenario is observed: 

(i) Above a crossover temperature \Tx{} 
the susceptibility minimum is well described by the asymptotic laws of MCT, 
allowing to identify consistently the critical temperature \Tc{} $\lesssim$ \Tx{}. 
For describing the full spectrum, including the $\alpha$-relaxation, 
it is sufficient to fix a single width parameter, i.\,e.,  $\beta_{\mathrm{CD}}= b$, which defines 
the high-frequency power-law behavior of $\chi''_\alpha (\nu)$. 
This has also been observed in toluene \cite{ref8}. 
Quite surprisingly these asymptotic predictions hold up to the fluid regime. 
No pecularities are observed around the melting temperature, $T_{\mathrm{m}} = 206\,\mathrm{K}$. 
(ii) Below \Tx{} the positions of the susceptibility minima do not follow the MCT predictions. 
Instead they are controlled by the appearance of the excess wing. 
The wing is believed to be a kind of precursor of the $\alpha$-process 
and/or a secondary relaxation process \cite{ref14,ref15}. 
Thus, we think including a hopping mechanism is not sufficient 
to understand the evolution of the susceptibility minimum below \Tx. 
(iii) Within a phenomenological approach one is able to single out the fast dynamics contribution. 
Power-law contributions are found with an exponent being constant above \Tx{} 
but increasing upon further cooling. 
This we take as a indication that the susceptibility crosses over to a white noise spectrum 
similarly as predicted by MCT. 
This trend is impossible to follow to a temperature at which $a' = 1$ holds, 
since the glass transition at \Tg{} interferes, 
resulting in a temperature independent exponent $a'$ at $T<T_{\mathrm{g}}$. 
(iv) The spectra of the fast dynamics allow to determine the non-ergodicity parameter, 
which clearly displays the anomaly predicted. 
In contrast to what has been reported from NS experiments 
a further change of its temperature dependence is observed at \Tg{},
i.\,e., the evolution of the fast dynamics is altered when the supercooled liquid's structure freezes. 
The discrepancy of LS and NS data may be explained by the fact
that the usual determination of $f(T)$ by NS includes also harmonic contributions 
and, as a consequence, the change of $f(T)$ around \Tg{} (and \Tc) is smeared out. 

We conclude that the fast dynamics in picoline is well described by idealized MCT 
even well above \Tc{} and 
even below the crossover temperature \Tx{}. 
The crossover to a white noise spectrum is ultimately related to the prevalence of the 
anomaly of $f(T)$. 
Why actually no ``knee'' appears 
and why the crossover to ``white noise'' is rather observed as a function of temperature 
than as a function of frequency 
is an open question. 
Finally, we note that the crossover temperature \Tx{}, 
where first deviations from the MCT predictions appear, 
is somehow higher than \Tc{} as obtained from extrapolating the scaling laws. 
This may explain certain discrepancies in identifying \Tc{} found in the literature.


\begin{thebibliography}{19}
\bibitem{ref1} W. G\"otze, \textit{J. Phys. Condens. Matter} \textbf{11}, A1 (1999).
\bibitem{ref2}	F. Mezei, M. Russina, \textit{J. Phys. Condens. Matter}, \textbf{11}, A341 (1999)
\bibitem{ref3}	W. G\"otze, Th. Voigtmann, \textit{Phys. Rev. E} \textbf{61}, 4133 (2000) 
\bibitem{ref4}	H. Z. Cummins, W.M. Du, G. Li, M. Fuchs, W. G\"otze, S. Hildebrand, A. Latz, G. Li, N.J. Tao, \textit{Phys. Rev. E}, \textbf{47}, 4223 (1993)
\bibitem{ref5}	N. Surovtsev, J. Wiedersich, V.N. Novikov, E. R\"ossler, \textit{Phys. Rev. B} \textbf{58}, 14888 (1998);
J. Wiedersich, PhD-thesis, Universit\"at Bayreuth (2000)
\bibitem{ref6}	J. Gapinski, W. Steffen, A. Patkowski, A.P. Sokolov, A. Kisliuk, U. Buchenau, 
M. Russina, F. Mezei, H. Schober, \textit{J. Chem. Phys.} \textbf{110}, 2312 (1999)
\bibitem{ref7}	H.C. Barshilia, G. Li, G.Q. Shen, H.Z. Cummins, \textit{Phys. Rev. E} \textbf{59}, 5625 (1999)
\bibitem{ref8}	J. Wiedersich, N. Surovtsev, E. R\"ossler, \textit{J. Chem. Phys.} \textbf{113}, 1143 (2000)
\bibitem{ref9}	A. Kudlik, S. Benkhof, T. Blochowicz, C. Tschirwitz, E. R\"ossler, 
\textit{J. Molec. Struct.} \textbf{479}, 201 (1999)
\bibitem{ref10}	 Ch. Tschirwitz, C. Alba-Simionesco, unpublished data
\bibitem{ref11}	 P. Lunkenheimer, U. Schneider, R. Brand, A. Loidl, 
\textit{Contemp. Phys.} \textbf{41}, 15 (2000)
\bibitem{ref12}	 P. K. Dixon, L. Wu, S. R. Nagel, B.D. Williams, J.P. Carini, 
\textit{Phys. Rev. Lett.} \textbf{65}, 1108 (1990) 
\bibitem{ref13}	 A. Kisliuk, V.N. Novikov, A.P. Sokolov,  
\textit{J. Polym. Sci., Polymer Physics} (2001) submitted
\bibitem{ref14}	 T. Blochowicz, A. Kudlik, S. Benkhof, J. Senker, E. R\"ossler, G. Hinze, 
\textit{J. Chem. Phys.} \textbf{110}, 12011 (1999)
\bibitem{ref15}	 U. Schneider, R. Brand, P. Lunkenheimer, A. Loidl, 
\textit{Phys. Rev. Lett.} \textbf{84}, 5560 (2000)
\end{thebibliography}
\end{document}